# Non-redundant optical phased array

TAICHIRO FUKUI,[1,*] RYOTA TANOMURA,[1] KENTO KOMATSU,[1] DAIJI YAMASHITA,[1] SHUN TAKAHASHI,[1] YOSHIAKI NAKANO,[1] AND TAKUO TANEMURA[1,2]

[1]*School of Engineering, The University of Tokyo, 7-3-1 Hongo, Bunkyo, Tokyo, 113-8656, Japan.*
[2]*tanemura@ee.t.u-tokyo.ac.jp*
*\*fukui@hotaka.t.u-tokyo.ac.jp*

**Abstract:** Optical phased array (OPA) is a promising beam-steering device for various applications such as light detection and ranging (LiDAR), optical projection, free-space optical communication and switching. However, the previously reported OPAs suffer from either an insufficiently small number of resolvable points, or a complicated control requirement due to an extremely large number of phase shifters. This work introduces a novel array configuration for OPA devices based on the non-redundant array (NRA) concept. Based on this design, we can realize high-resolution OPA whose number of resolvable points scales with $N^2$. In contrast, that of traditional OPAs scales only with $N$. Thus, a significant reduction in the number of required phase shifters can be attained without sacrificing the number of resolvable points. We first investigate the impact of employing the NRA theoretically by considering the autocorrelation function of the array layout. We then develop a Costas-array-based silicon OPA and experimentally demonstrate 2D beam steering with ~19,000 resolvable points using only 127 phase shifters. This corresponds to the largest number of resolvable points achieved by an OPA without sweeping the wavelength to the best of our knowledge.

## 1. Introduction

Optical phased array (OPA) [1–31] is a promising non-mechanical beam-steering device for various applications, such as light detection and ranging (LiDAR) [8, 9], free-space optical communication [9,10], photonic switching [11–14], optical projection [15,16], and remote sensing [17]. It can be monolithically integrated with an electrical driver circuit on a CMOS compatible silicon photonic chip [18-20]. Moreover, by employing III-V/Si hybrid or heterogeneous integration platforms, light sources and optical amplifiers may also be mounted on the same chip to realize an extremely compact and low-cost device [21-23]. The operation speed of OPAs can exceed hundreds of MHz by employing electro-optic phase shifters [24-26]. On the other hand, by selecting appropriate photonic integration platforms, OPAs can also operate at versatile wavelength ranges, including the visible [16], near-infrared [27, 28], and mid-infrared [29, 30].

One of the inherent challenges of the OPA-based beam-steering devices is the difficulty in achieving a high spatial resolution without increasing the complexity. An OPA enables beam steering by emitting phase-controlled lightwaves from numerous optical antennas. In a traditional OPA with its optical antennas arranged in a uniform array, the number of resolvable points is nearly equal to the number of antennas $N$ [1]. Therefore, to realize an OPA that can resolve several thousand or more points, as required in many imaging-related applications, the array size should also be scaled to several thousands of elements [31].

As the array size increases, however, various issues, including the complexity and power consumption of the driver circuit to achieve accurate phase control, become prominent. In particular, the need for precise control of numerous optical phase shifters imposes a serious issue as $N$ increases. Due to the inevitable deviations of the waveguides across the array and crosstalk among the phase shifters [32,33], real-time monitoring and calibration of optical phases are mandatory in the practical systems. While this process can be significantly simplified by the speckle-based imaging methods [17, 34–37], such schemes may not be employed in applications that require focused optical beams. In most of the laboratory experiments, therefore, the far-field pattern (FFP) from the OPA is monitored by an external camera, which is then used to calibrate all phase shifters through an iterative optimization process [6-9,15,16,18-24]. More recently, several on-chip phase calibration methods have been demonstrated [38-40], aiming for commercial applications of OPA-based beam-steering devices. As a result, additional on-chip optical interferometers and photodetectors as well as feedback control circuits are required, which would introduce extra complexity and device footprint in proportion to $N$. It is, therefore, essential to reduce the number of phase shifters $N$ without sacrificing the spatial resolution.

One method to alleviate this issue is placing the optical antennas sparsely with non-uniform spacings [16,20,23,41-44], so that the number of resolvable points can be enhanced without increasing $N$ significantly. In the previous works, several classes of non-uniform arrays have been applied to OPAs, such as the Gaussian arrays [23], pseudo-random arrays [41], and iteratively optimized arrays [16,20]. For example, 500 resolvable points were obtained with 128 phase shifters in [41], corresponding to nearly five-fold improvement from the uniform array with a same number of phase shifters. On the other hand, the theoretical limit of spatial resolution of such non-uniform OPAs has not been clarified. Moreover, none of the previous work has demonstrated extremely high-resolution beam steering with more than 10,000 resolvable points using a single wavelength.

Here, we elucidate the fundamental limit of the achievable spatial resolution of an OPA and demonstrate that ultrahigh-resolution two-dimensional (2D) beam steering with up to $N^2$ resolvable points can be obtained by employing the concept of non-redundant array (NRA). We theoretically derive that the spatial resolution of OPA is directly linked to the autocorrelation of the antenna locations, from which the maximum number of resolvable points is shown to scale as $N^2$. We then propose that this limit can be approached by employing an NRA, which exhibits a nearly optimal uniformly distributed autocorrelation. To experimentally demonstrate this concept, we fabricate a silicon photonic OPA chip with 127 phase shifters, whose output grating antennas are spaced based on the Costas array. Using the fabricated chip, we demonstrate extremely high-resolution 2D beam steering; a beam with a full-width at half-maximum (FWHM) of 0.0428° is generated, which is steered across 5.92° × 5.92° field of view (FOV). From these results, the number of resolvable points by our OPA is derived to be ~19,100, corresponding to the theoretical limit of ~$N^2$. To the best of our knowledge, this is the largest number of resolvable points achieved by an OPA operating at a single wavelength.

## 2. Device concept

### 2.1 Theory

Figure 1 schematically shows the 2D OPA demonstrated in this work. The optical antennas are placed sparsely with non-uniform spacings. We will show that the locations of these antennas can be selected judiciously to eliminate the redundancy in achieving high-resolution beam steering.

We first derive the general relationship between the intensity distribution of the FFP and the near field pattern (NFP) of an OPA with $N$ waveguides. Figure 2 describes the definitions of the electric field distributions $E(\mathbf{r})$ and $F(\boldsymbol{\xi})$ at the NFP and FFP planes, respectively. Here, $\mathbf{r}=(x,y)$ is the coordinate at the NFP plane, while $\boldsymbol{\xi}=(\theta,\psi)$ denote the lateral and longitudinal angles at the FFP. We can express $E(\mathbf{r})$ as

$$E(\mathbf{r}) = \sum_{n=1}^{N} C_n u(\mathbf{r}-\mathbf{r}_n) = u(\mathbf{r}) * \left[\sum_{n=1}^{N} C_n \delta(\mathbf{r}-\mathbf{r}_n)\right] \equiv u(\mathbf{r}) * a(\mathbf{r}), \qquad (1)$$

where $u(\mathbf{r})$ describes the electric field distribution from a single optical antenna at the vicinity of the emission plane, $f * g \equiv \iint f(\mathbf{r}-\mathbf{r}')g(\mathbf{r}')\mathrm{d}^2\mathbf{r}'$ denotes the convolution operation, $\mathbf{r}_n = (x_n, y_n)$ represents the position of the $n^{\text{th}}$ $(n=1,2,...,N)$ optical antenna, $\delta(\mathbf{r})$ is the Dirac's delta function, and $C_n = |C_n|\exp(i\phi_n)$ is the complex amplitude of the light emitted from $n^{\text{th}}$ optical antenna. For convenience, we define $a(\mathbf{r}) \equiv \sum_{n=1}^{N} C_n \delta(\mathbf{r} - \mathbf{r}_n)$ to represent the entire array layout.

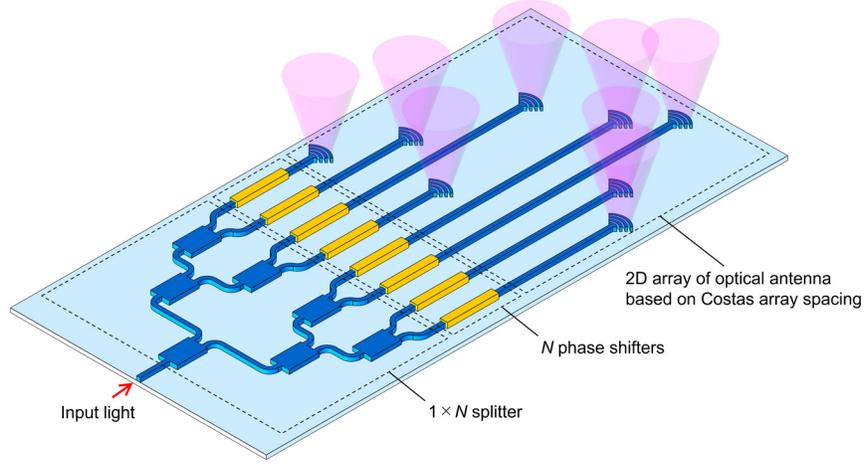

Fig. 1 Schematic of the 2D non-redundant OPA demonstrated in this work.

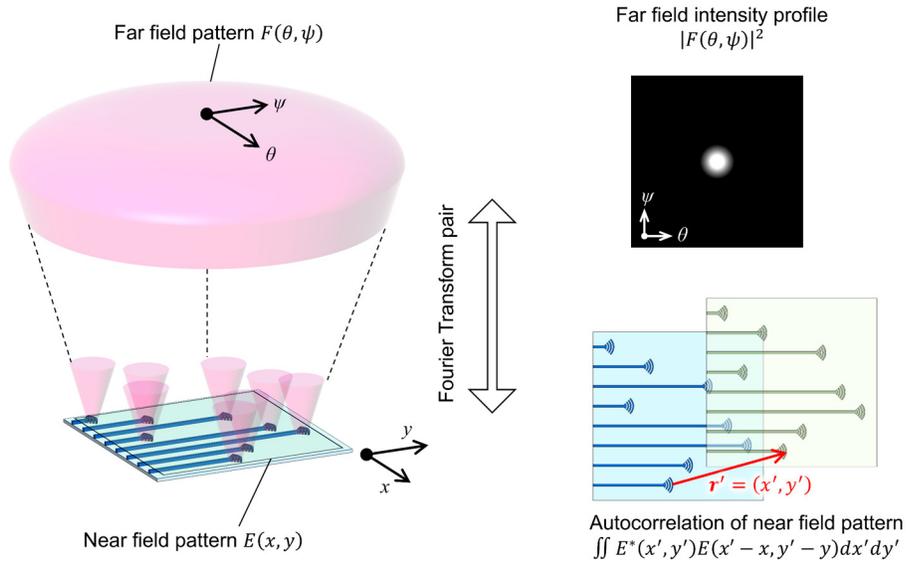

Fig. 2 The definitions of variables and functions. The intensity distribution of FFP is the Fourier transform of the autocorrelation function of NFP.

Since the FFP of the OPA is given as the Fourier transform of the NFP, $F(\boldsymbol{\xi})$ is written as

$$F(\boldsymbol{\xi}) = \iint E(\mathbf{r}) \exp\left[ik_0 (\mathbf{r} \cdot \boldsymbol{\xi})\right] d^2\mathbf{r}, \qquad (2)$$

where $k_0$ is the wavenumber of the light in the vacuum. By inserting Eq. (1) to Eq. (2), we obtain

$$F(\xi) = U(\xi)\sum_{n=1}^{N} C_n \exp[ik_0(\mathbf{r}_n \cdot \xi)]. \tag{3}$$

Here, $U(\xi)$ is the Fourier transform of $u(\mathbf{r})$, i.e.,

$$U(\xi) = \iint u(\mathbf{r})\exp[ik_0(\mathbf{r}\cdot\xi)]d^2\mathbf{r}, \tag{4}$$

describing the FFP of the electric field from a single optical antenna. Finally, the intensity distribution of the FFP, which is important for the beam-steering applications, can be expressed as

$$I(\xi) \equiv |F(\xi)|^2 = |U(\xi)|^2 \sum_{n=1}^{N}\sum_{m=1}^{N} C_n C_m^* \exp[ik_0(\mathbf{r}_n - \mathbf{r}_m)\cdot\xi]. \tag{5}$$

In Eq. (5), each term inside the summation, $C_n C_m^* |U(\xi)|^2 \exp[ik_0(\mathbf{r}_n - \mathbf{r}_m)\cdot\xi]$, physically represents the interference pattern generated by the lightwaves from $m^{th}$ port and $n^{th}$ port. When $\phi_1 = \phi_2 = ... = \phi_N$, a strong intensity peak is obtained at $\xi = (0,0)$. This corresponds to the case of forming an equal phase plane at the emitter plane. The intensity peak can then be steered to an arbitrary angle $\xi_0 = (\theta_0, \psi_0)$ within the FOV by tuning the optical phases $\phi_n$ so that the condition

$$k_0(\mathbf{r}_n - \mathbf{r}_m)\cdot\xi + \phi_n - \phi_m = k_0(\mathbf{r}_n - \mathbf{r}_m)\cdot(\xi - \xi_0) \tag{6}$$

is satisfied for all $n$ and $m$ $(=1,2,...,N)$. Note that this case corresponds to a situation of forming an equal phase plane tilted by an angle $\xi_0$.

Equation (5) provides an important insight that the displacement vectors $\Delta_{n,m} \equiv \mathbf{r}_n - \mathbf{r}_m$ between $N$ optical antennas defines the spatial frequency spectrum of the FFP, and, therefore, have a critical impact on determining the FFP. To investigate this aspect quantitatively, we now derive a different expression of $I(\xi)$. From Eq. (2),

$$\begin{aligned} I(\xi) = |F(\xi)|^2 &= \iint\iint E(\mathbf{r})E^*(\mathbf{r}')\exp[ik_0(\mathbf{r}-\mathbf{r}')\cdot\xi]d^2\mathbf{r}d^2\mathbf{r}' \\ &= \iint\iint E(\mathbf{r}'+\Delta)E^*(\mathbf{r}')\exp[ik_0(\Delta\cdot\xi)]d^2\Delta d^2\mathbf{r}' \\ &= \iint R_{EE}(\Delta)\exp[ik_0(\Delta\cdot\xi)]d^2\Delta \end{aligned} \tag{7}$$

where $\Delta \equiv \mathbf{r} - \mathbf{r}'$. We define $R_{EE}(\Delta)$ as

$$R_{EE}(\Delta) \equiv \iint E(\mathbf{r}+\Delta)E^*(\mathbf{r})d^2\mathbf{r}, \tag{8}$$

which denotes the autocorrelation function of $E(\mathbf{r})$ in the spatial domain. Eq. (7) implies that $I(\xi)$ is the Fourier transform of $R_{EE}(\Delta)$. To obtain ideal $I(\xi)$ with a fine sharp beam profile, therefore, it is essential to obtain as flat and broad $R_{EE}(\Delta)$ as possible.

By substituting Eq. (1) to Eq. (8), we obtain

$$\begin{aligned} R_{EE}(\Delta) &= \sum_{n=1}^{N}\sum_{m=1}^{N} C_n C_m^* \iint u(\mathbf{r}+\Delta-\mathbf{r}_n)u(\mathbf{r}-\mathbf{r}_m)d^2\mathbf{r} \\ &= \sum_{n=1}^{N}\sum_{m=1}^{N} C_n C_m^* R_{uu}(\Delta - \Delta_{n,m}) \\ &= R_{uu}(\Delta) * \sum_{n=1}^{N}\sum_{m=1}^{N} C_n C_m^* \delta(\Delta - \Delta_{n,m}) \\ &= R_{uu}(\Delta) * R_{aa}(\Delta) \end{aligned} \tag{9}$$

Here we define

$$R_{uu}(\Delta) \equiv \iint u(\mathbf{r}+\Delta)u^*(\mathbf{r})d^2\mathbf{r}, \tag{10}$$

and

$$R_{aa}(\Delta) \equiv \iint a(\mathbf{r}+\Delta)a^*(\mathbf{r})d^2\mathbf{r} = \sum_{n=1}^{N}\sum_{m=1}^{N}C_n C_m^* \delta(\Delta - \Delta_{n,m}), \quad (11)$$

which are the autocorrelation functions of the electric field distribution from each antenna and the array layout, respectively.

From Eqs. (7) and (9), we can conclude that the spatial Fourier transform of FFP from an OPA is represented by $R_{EE}(\Delta)$, which is expressed as the convolution of $R_{uu}(\Delta)$ and $R_{aa}(\Delta)$. From Eq. (10), however, $R_{uu}(\Delta)$ quickly drops to zero when $\Delta$ exceeds the physical size of a grating antenna. As a result, the bandwidth of $R_{EE}(\Delta)$, or in other words, the maximal spatial frequency component of FFP, is determined purely by $R_{aa}(\Delta)$ [45]. Therefore, a broad and flat $R_{EE}(\Delta)$ is obtained if we could engineer the array layout to spread $R_{aa}(\Delta)$ as broad as possible. From the definition of $R_{aa}(\Delta)$ in Eq. (11), such condition is realized when every $\Delta_{n,m}$ are mutually different. In other words, all the interference patterns represented by $C_n C_m^* |U(\xi)|^2 \exp[ik_0(\Delta_{n,m}\cdot\xi)]$ in equation (5) should have different spatial frequencies. At the same time, $\Delta_{n,m}$ should be distributed as uniformly as possible so that the beam shape would not degrade.

Therefore, in the next section, we will only focus on $R_{aa}(\Delta)$ to clarify the impact of the array layout on the FFP. For simplicity, we also assume $C_1 = C_2 = ... = C_N = 1$, which corresponds to the case of focusing the beam at $\xi_0 = (0,0)$ [46]. Under these assumptions, $R_{aa}(\Delta)$ in Eq. (11) reduces to

$$R_{aa}(\Delta) = \sum_{n=1}^{N}\sum_{m=1}^{N}\delta(\Delta - \Delta_{n,m}). \quad (12)$$

## 2.2 Concept of non-redundant OPA

In the proposed non-redundant OPA shown in Fig. 1, the antennas are located at judiciously selected locations, so that $R_{aa}(\Delta)$ in Eq. (12) exhibits a broad and flat distribution. We should note that such concept has been recognized as NRA in the fields of radio astronomy [47] as well as sound navigation and ranging (SONAR) [48] and has been proved to be an effective means to maximize the efficiency of sampling the spatial frequency with minimal resources. Here, we apply this concept to the OPA for the first time.

For the simplicity of explanation, we first consider a 1D OPA, where the optical antennas are located along the *x* axis (with constant *y*) in Fig. 2. The most common example of 1D NRA is known as the Golomb ruler [49]. It defines a set of integer positions, which satisfies a condition that the distance between any pair of these positions is mutually distinct. The shortest Golomb ruler for a given number of element *N* is called an optimal Golomb ruler, which is suitable to realize 1D OPA since it maximizes the degree of uniformness of the autocorrelation. As example cases, we compare $a(x)$ of a 1D uniform array (*N* = 14) [Fig. 3(a)] with an optimal Golomb ruler for *N* = 14 [Fig. 3(b)]. Both arrays are drawn with the same dimensionless unit length (i.e., minimal spacing) of 1 along the *x* direction. In Figs. 3(c) and (d), we compare the autocorrelation functions $R_{aa}(\Delta)$ for both cases. As shown in Fig. 3(c), in case of the uniform array, due to the redundancy (i.e., the duplication of $\Delta_{n,m}$), $R_{aa}(\Delta)$ exhibits a narrow triangular shape with the width of $\Delta x = 26$ ($= 2N - 2$). In contrast, we see in Fig. 3(d) that the $\mathrm{R}_{aa}(\boldsymbol{\Delta})$ of the optimal Golomb ruler is distributed uniformly over a large range of $\Delta x = 254$ ($\sim N^2$). Here, the sharp spike at $\Delta x = 0$ corresponds to the inevitable DC element, which originates from the term $\sum_{n=1}^{N}\delta(\Delta - \Delta_{n,n}) = \sum_{n=1}^{N}\delta(\Delta)$ in Eq. (12).

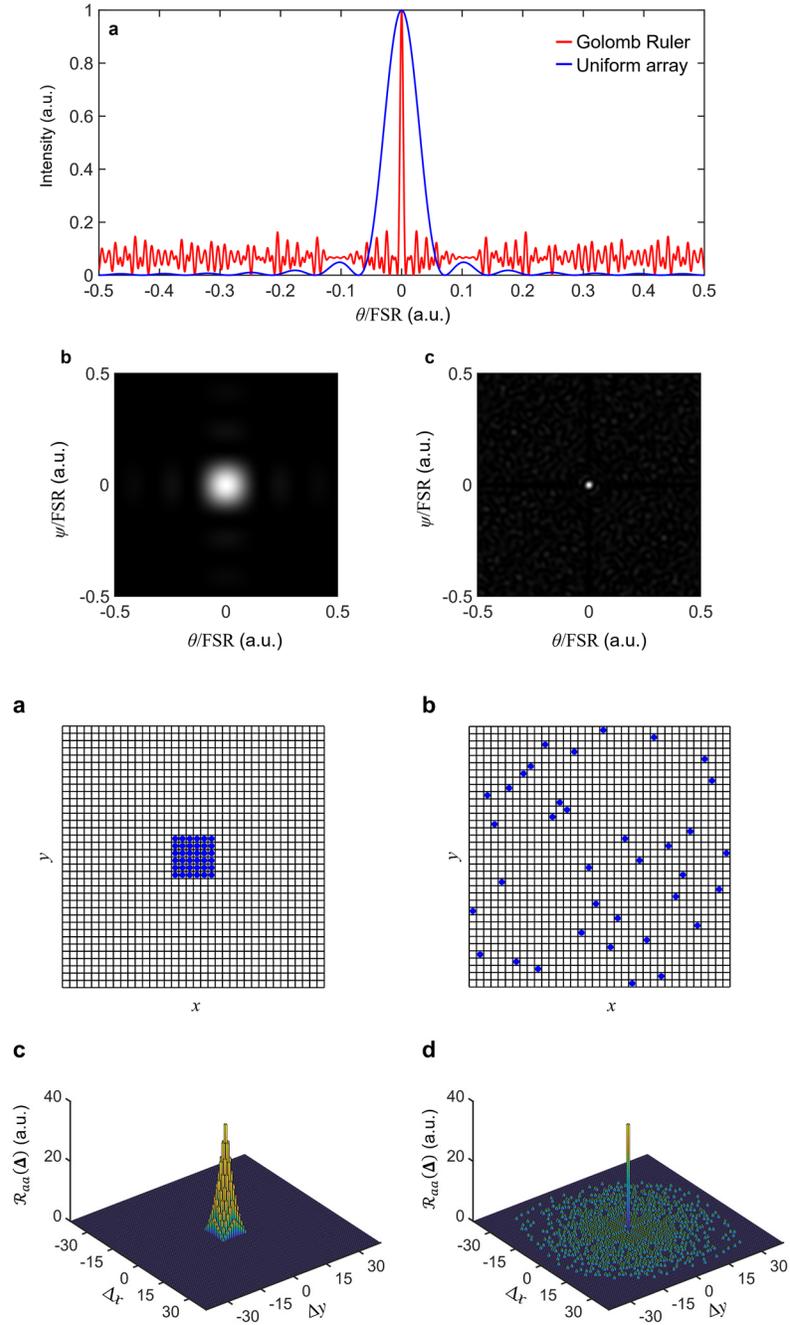

Fig. 4. Comparison of a 2D uniform array and a 2D NRA ($N = 36$). (a) $a(\mathbf{r})$ of 2D uniform array. (b) $a(\mathbf{r})$ of 2D NRA (Costas array). (c) $\mathrm{R}_{aa}(\boldsymbol{\Delta})$ of 2D uniform array. (d) $\mathrm{R}_{aa}(\boldsymbol{\Delta})$ of 2D NRA.

Since the maximum number of mutually different $\Delta_{n,m}$ in Eq. (12) is $N^2 - N + 1 = 2\,_N C_2 + 1$ (number of cases for choosing two from $N$ elements while distinguishing their order, and an additional 1 for the DC element), $R_{aa}(\Delta)$ can be as broad as $N^2 - N + 1$ with a uniform density. Note that it may be slightly broader (i.e., the autocorrelation width may exceed $N^2 - N + 1$) when $R_{aa}(\Delta)$ is not perfectly uniform at the cost of some penalty in the beam quality. This limit indeed coincides with the maximum number of independent intensity patterns that can be generated by $N$ phase shifters [50], implying that NRA corresponds to an optimal antenna configuration of the OPA-based imaging. It is worth noting that the number of resolvable points of an NRA-based OPA scales by $N^2$ since the beam width is inversely proportional to the bandwidth of the spatial frequency spectrum. This feature is in clear contrast with the case of uniformly spaced OPA, where the number of resolvable points scales only by $N$.

We now consider a 2D OPA with NRA. While there are several classes of 2D NRA, we employ the Costas array as a typical example. Costas array is an 2D array, where $N$ elements are placed on an $N \times N$ grid in a way such that each row and column contains only one element. Similar to the 1D Golomb ruler, the 2D displacement vectors $\Delta_{n,m}$ between all pairs of elements in a Costas array are mutually distinct. NRAs have been studied extensively in the field of mathematics and NRAs with $N > 100$ have been discovered to date [51-54].

In Fig. 4, we compare a 2D uniform array and a Costas array with $N = 36$. In case of 2D uniform array, the antennas are located in a 6×6 uniform grid [Fig. 4(a)]. In contrast, we can see in Fig. 4(b) that the antennas in the Costas array are located sparsely in a 36×36 uniform grid. Again, in the case of a uniform array described in Fig. 4(a), $R_{aa}(\Delta)$ is narrow and pyramid-like shaped as shown in Fig. 4(c), which fits within a range of $|\Delta x| < 5$, $|\Delta y| < 5$ [corresponding to $(2\sqrt{N} - 2) \times (2\sqrt{N} - 2)$]. On the other hand, the Costas array [Fig. 4(b)] has a broad and flat $R_{aa}(\Delta)$ [Fig. 4(d)], which distributes over a large range of $|\Delta x| < 35$, $|\Delta y| < 35$ [corresponding to $(2N - 2) \times (2N - 2)$].

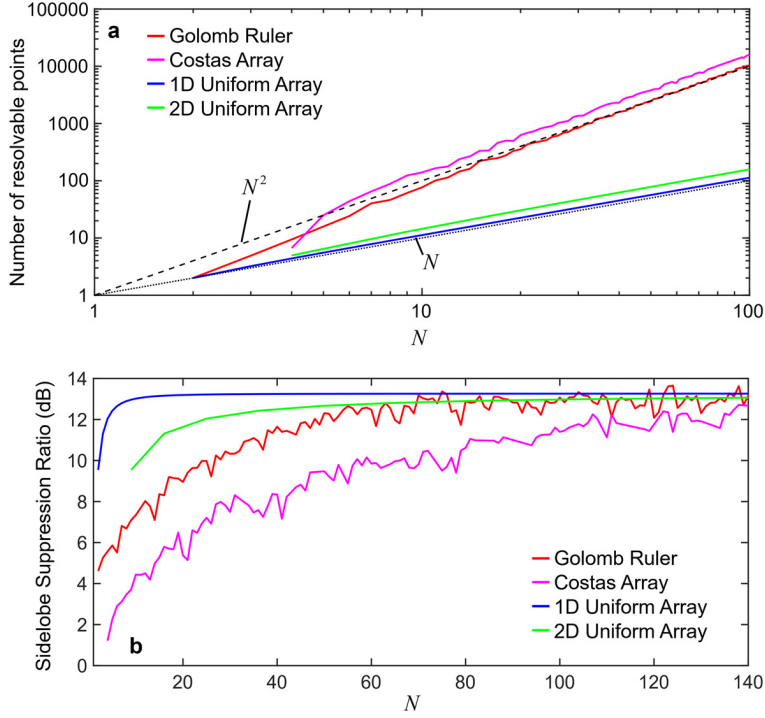

Fig. 6 Scalabilities of Golomb-ruler-based OPA (red), Costas-array-based OPA (pink), 1D uniform OPA (blue), and 2D uniform OPA (green). (a) $N$ dependency of the number of resolvable points. As a reference, $N$ (black dotted) and $N^2$ (black broken) are also plotted. (b) $N$ dependency of the sidelobe suppression ratio.

Next, we numerically derive the FFP of the OPA to directly investigate the impact of employing the NRA. In Fig. 5(a), we compare the FFPs of 1D OPAs with $N = 14$ based on the uniform array (blue) and the NRA (red) under a condition when all phases are aligned. We can clearly observe that the FFP of the NRA-based-OPA can be much narrower compared to that of the uniform OPA. Quantitatively, the number of the resolvable points [defined by FOV/FWHM] is derived to be 16 points ($\sim N$) for the uniform OPA, while it is enhanced to 161 points ($\sim N^2$) for the NRA-based OPA. Similarly, we show the FFPs of 2D OPAs based on the uniform array [Fig. 5(b)] and the Costas array [Fig. 5(c)] with $N = 36$. Again, we can see that larger number of resolvable points [defined as (full FOV)/(focused beam size), where the beam size is derived as the area with the intensity stronger than the half of the peak intensity] can be obtained by employing the NRA [2,063 points ($\sim N^2$)] compared to the uniform array [56 points ($\sim N$)].

Finally, the scalability of the NRA-based OPA is investigated by analyzing the FFP with increasing $N$. Figure 6(a) shows how the number of resolvable points scales with $N$. While the number of resolvable points scales with $N$ (black-dotted) for the uniform OPAs (1D: blue, 2D: green), those of the NRA-based OPAs (Golomb ruler: red, Costas array: pink) scale with $N^2$

(black-broken). Figure 6(b) plots the sidelobe suppression ratio as a function of $N$. Although the sidelobe suppression is degraded for the NRA-based OPA at small $N$, it soon exceeds 10 dB and approaches that of uniform OPAs at a moderate $N$ of around 80.

## 3. Experimental Result

In order to experimentally demonstrate the concept of NRA-based-OPA, we fabricated a silicon photonic OPA chip with $N = 127$ using an 8-inch silicon-on-insulator (SOI) multi-project wafer foundry service. The SOI wafer consisted of 220 nm silicon layer and 2 μm buried oxide (BOX) layer. The optical microscope image of the device is shown in Fig. 7(a). We employed 7-stages of cascaded 1×2 multimode interference (MMI) couplers to distribute the input light to 128 waveguides (while one of them was used as an alignment port). Each of the waveguides was equipped with a 220-μm-long TiN thermo-optic phase shifter [Fig. 7(b)] and a 10-μm-long grating emitter [Fig. 7(c)]. These grating emitters were placed at a Costas array (acquired from [50]) to generate a high-resolution beam. The waveguides were carefully routed so that the optical path lengths of all 127 ports were identical. The unit length of the Costas array spacing was set to be 15 μm, from which the FOV was derived to be 5.92° × 5.92°. The fabricated OPA chip was mounted on an AlN chip carrier and electrically connected to a driver circuit.

We evaluated the beam-steering properties of the OPA chip using the system depicted in Fig. 8. A 1550-nm wavelength continuous-wave light from a laser source was aligned to the transverse-electric (TE) mode using a polarization controller (PC) and was coupled to the OPA chip via a lensed fiber. The output light from the OPA chip was relayed through a 4-f system consisted of two plano-convex lenses (focal lengths: $f_1 = f_2 = 100$ mm. diameters: $\phi_1 = \phi_2 = 50$ mm). We used the near field pattern (NFP) of the chip formed after the 4-f system for the optical alignment and to filter undesired light using a mechanical slit. We cascaded another 2-f system consisted of a plano-convex lens ($f_3$ = 100 mm, $\phi_3$ = 50 mm) to obtain the FFP from the OPA chip. The FFP of the chip was observed by an InGaAs camera (Hamamatsu Photonics C12741-03). We employed a field-programmable gate array (FPGA) driver circuit with 10-bit digital-to-analog converters (DAC) to drive all 127 phase shifters on the OPA independently. An iterative optimization algorithm based on the rotating element electric field vector method [55,56] was employed to derive the optimal driving condition to obtain a desired FFP. Prior to the beam evaluation, we applied random patterns to the OPA and obtained the average of the FFP intensity to acquire the envelope function of the FFP. Then, the measured beam profiles were calibrated to eliminate the effect of the envelope function, so that the impact of the array structure was purely evaluated.

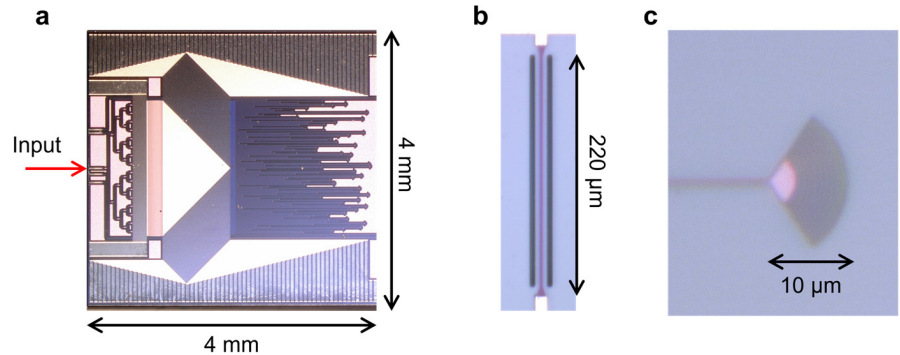

Fig. 7 Optical microscope images of the Si OPA chip. (a) The entire device. (b) TiN thermo-optic phase shifter. (c) Grating emitter.

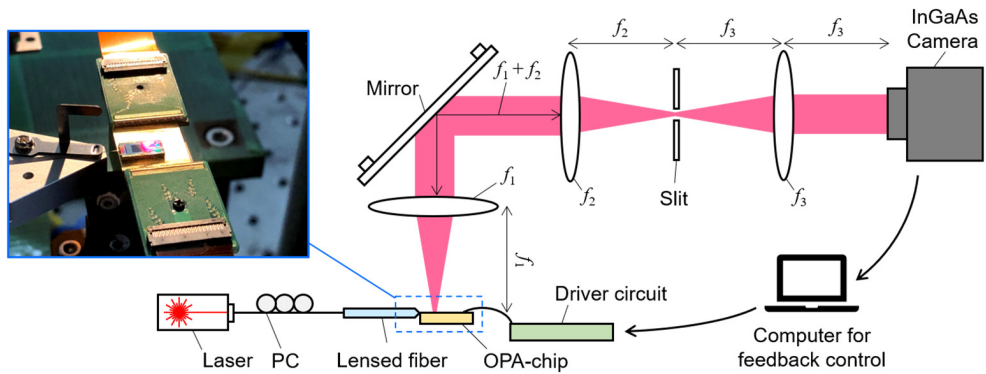

Fig. 8 Schematic of the experiment system.

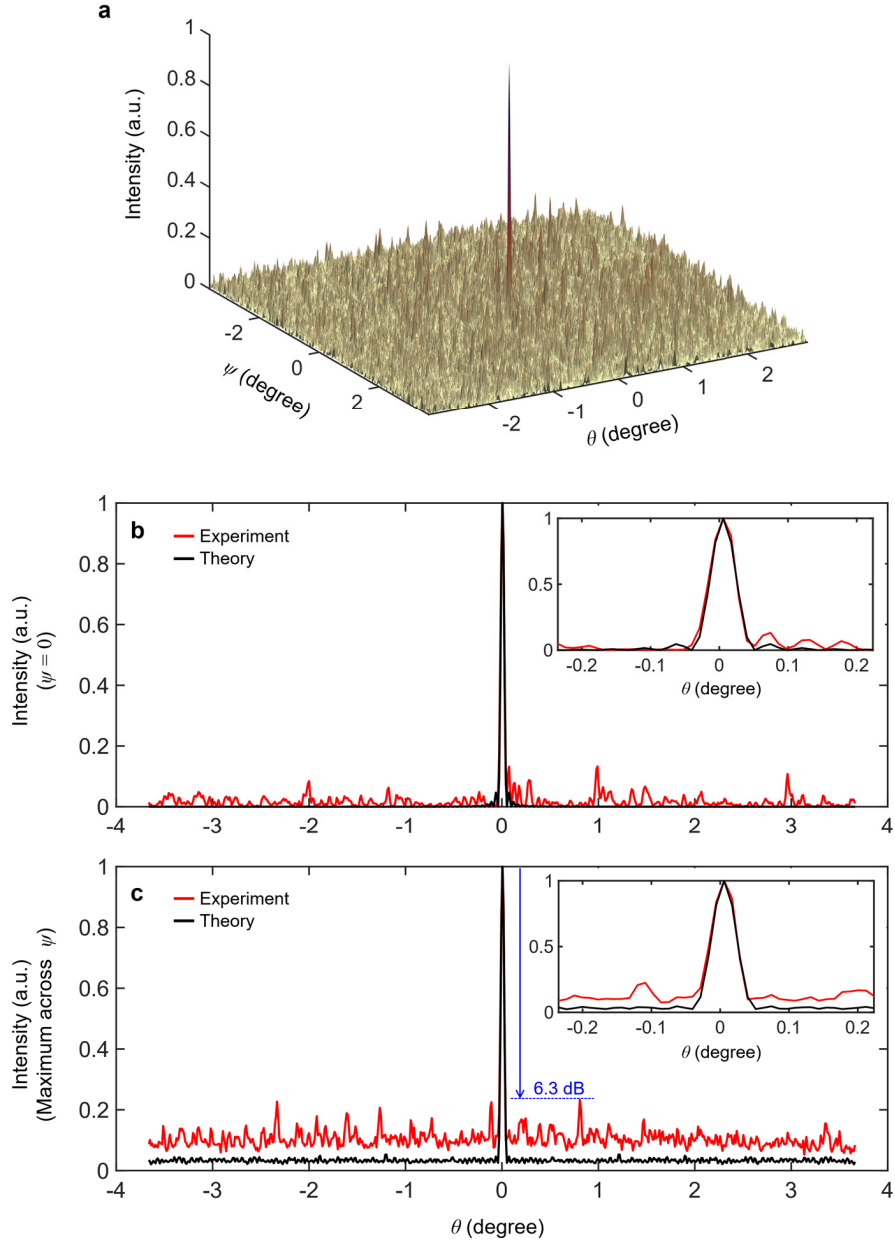

Fig. 9 Measured beam intensity profiles $I(\xi)$ after optimization. (a) Measured overall beam profile, emitted from the OPA chip. (b) The cross section of the beam at $\psi = 0$ [i.e., $I(\theta, \psi = 0)$] and (c) the maximum value along the $\psi$ direction $\{f(\theta) = \max[I(\theta, \psi); \psi \text{ in FOV}]\}$ for each $\theta$. In (b) and (c), magnified views are shown in the insets. The experimental results (red) are compared with the theoretical profiles (black).

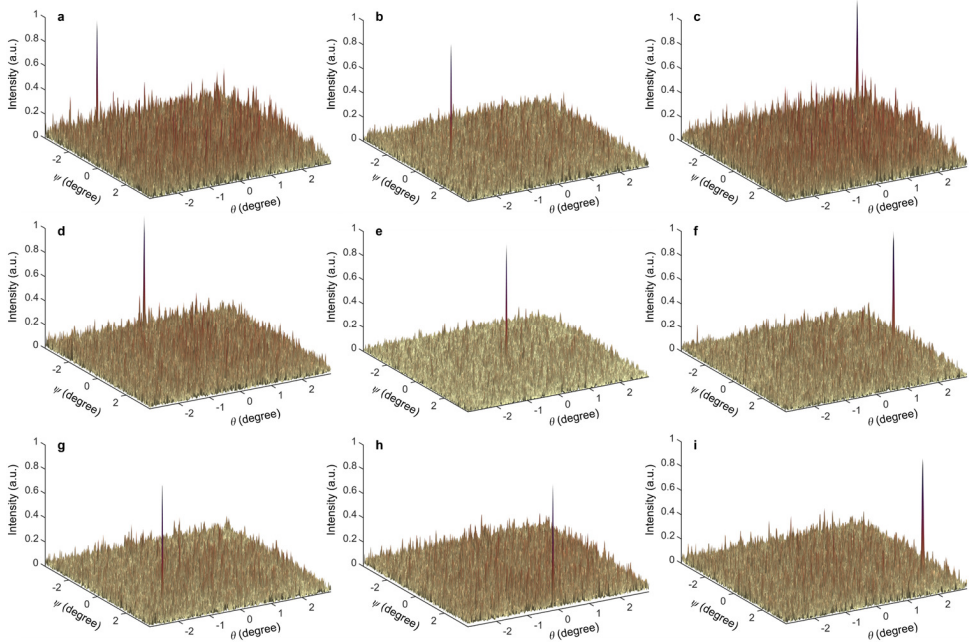

Fig. 10 The experimental results of beam steering to 9 points. (a) $(\theta, \psi) = (-2.52°, 2.22°)$, FWHM = 0.0489°. (b) $(\theta, \psi) = (0°, 2.22°)$, FWHM = 0.0457°. (c) $(\theta, \psi) = (2.06°, 2.22°)$, FWHM = 0.0514°. (d) $(\theta, \psi) = (-2.53°, 0°)$, FWHM = 0.0423°. (e) $(\theta, \psi) = (0°, 0°)$, FWHM = 0.0428°. (f) $(\theta, \psi) = (3.2°, 0°)$, FWHM = 0.044°. (g) $(\theta, \psi) = (-2.52°, -1.79°)$, FWHM = 0.0488°. (h) $(\theta, \psi) = (0°, -1.79°)$, FWHM = 0.0340°. (i) $(\theta, \psi) = (2.06°, -1.79°)$, FWHM = 0.0423°.

An example of the measured beam profile $I(\xi)$ after the optimization is shown in Fig. 9(a). A fine beam with a sharp intensity peak at $(\theta, \psi) = (0°, 0°)$ is obtained. The FWHM (defined by the average of the FWHM in $\theta$ and $\psi$ directions) of the beam is 0.0428°. The cross-section of the beam at the center ($\psi = 0$) is illustrated in Fig. 9(b), where the experimental result (red) is compared with the theoretically obtained profile (black). Excellent agreement between the experiment and the theory is obtained. Fig. 9(c) shows the maximum intensity along $\psi$ as a function of $\theta$, from which the sidelobe suppression ratio is derived to be 6.3 dB. Theoretically, this can be further reduced to 12.3 dB.

Figure 10 shows the results of 2D beam steering over the 5.92° × 5.92°FOV. The FWHM of the beam is roughly constant over the entire FOV in agreement with the theory presented in Section 2. The residual deviation is attributed to the nonzero aberrations at the edges of the lens system. For clarity, we illustrate in Fig. 11 the measured beam steering result in the $\theta$ direction [Fig. 11(a)] and the $\psi$ direction [Fig. 11(b)] at $\psi = 0°$ and $\theta = 0°$, respectively. Beam steering across the entire FSR is successfully demonstrated in both axes. From these results, we can derive the number of resolvable points as (FOV/FWHM)$^2$ ~19,100 points, corresponding to the theoretical limit of ~$N^2$. To the best of our knowledge, this is the largest number of resolvable points achieved by an OPA without sweeping the wavelength.

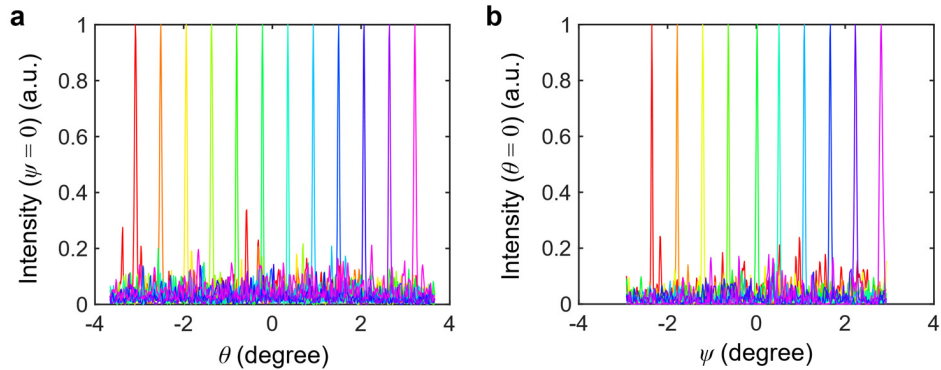

Fig. 11 Measured beam steering result (a) across $\theta$ ($\psi = 0$), and (b) across $\psi$ ($\theta = 0$).

## 4. Conclusion

We have proposed and demonstrated an ultra-high-resolution beam-steering OPA device based on the NRA concept. The redundancy of the array is eliminated by avoiding duplication of the displacement vectors $\Delta_{n,m}$ between arbitrary pairs of the antenna elements. As a result, the number of resolvable points scales with $N^2$, in clear contrast to the conventional uniform OPA, whose spatial resolution scales only linearly with $N$. To verify this concept, we fabricated a silicon photonic 2D OPA with $N = 127$ based on the optimized NRA design of Costas array. We experimentally demonstrated 2D beam steering with approximately ~19,100 resolvable points, which is, to our knowledge, the largest number of resolvable points achieved by an OPA without sweeping the wavelength. The demonstrated capability of realizing extremely high resolution with a minimal number of phase shifters paves the way towards applying OPA technology in versatile fields, including optical sensing, switching, and communication.

## Funding

Japan Society for the Promotion of Sciences (18H03769, 21J11982)).

## Acknowledgments

The authors thank Keiichi Iguchi for the development of the driver circuit. The authors thank Yasuyuki Ozeki and Yusuke Kohno for fruitful discussions.

## Disclosures

The authors declare no conflicts of interest.

## References

1. M. J. R. Heck, "Highly integrated optical phased arrays: photonic integrated circuits for optical beam shaping and beam steering," Nanophotonics **6**(1), 93–107 (2017).
2. K. Van Acoleyen, W. Bogaerts, J. Jágerská, N. Le Thomas, R. Houdré, and R. Baets, "Off-chip beam steering with a one-dimensional optical phased array on silicon-on-insulator," Opt. Lett. **34**(9), 1477–1479 (2009).
3. K. Van Acoleyen, H. Rogier, and R. Baets, "Two-dimensional optical phased array antenna on silicon-on-insulator," Opt. Express **18**(13), 13655–13660 (2010).

4. J. K. Doylend, M. J. R. Heck, J. T. Bovington, J. D. Peters, L. A. Coldren, and J. E. Bowers, "Two-dimensional free-space beam steering with an optical phased array on silicon-on-insulator," Opt. Express **19**(22), 21595–21604 (2011).
5. J. Sun, E. Timurdogan, A. Yaacobi, E. S. Hosseini, and M. R.Watts, "Large-scale nanophotonic phased array," Nature **493**, 195–199 (2013).
6. D. Kwong, A. Hosseini, J. Covey, Y. Zhang, X. Xu, H. Subbaraman, and R. T. Chen, "On-chip silicon optical phased array for two-dimensional beam steering," Opt. Lett. **39**(4), 941-944 (2014).
7. S. A. Miller, Y.-C. Chang, C. T. Phare, M. C. Shin, M. Zadka, S. P. Roberts, B. Stern, X. Ji, A. Mohanty, O. A. Jimenez Gordillo, U. D. Dave, and M. Lipson, "Large-scale optical phased array using a low-power multi-pass silicon photonic platform," Optica **7**(1), 3–6 (2020).
8. C. V. Poulton, A. Yaacobi, D. B. Cole, M. J. Byrd, M. Raval, D. Vermeulen, and M. R. Watts, "Coherent solid-state LIDAR with silicon photonic optical phased arrays," Opt. Lett. **42**, 4091–4094 (2017).
9. C. V. Poulton, M. J. Byrd, P. Russo, E. Timurdogan, M. Khandaker, D. Vermeulen, and M. R. Watts, "Long-range LiDAR and free-space data communication with high-performance optical phased arrays," IEEE J. Sel. Top. Quantum Electron. **25**(5), 7700108 (2019).
10. W. S. Rabinovich, P. G. Goetz, M. Pruessner, R. Mahon, M. S. Ferraro, D. Park, E. Fleet, M. J. DePrenger, "Free space optical communication link using a silicon photonic optical phased array," Proc. SPIE **9354**, 93540B (2015)
11. T. Tanemura, I. Soganci, T. Oyama, T. Ohyama , S. Mino , K. Williams, N. Calabretta, H. Dorren, and Y. Nakano, "Large-capacity compact optical buffer based on InP integrated phased-array switch and coiled fiber delay lines," J. Lightwave Technol. **29**(4), 396-402 (2011).
12. I. M. Soganci, T. Tanemura, K. Takeda, M. Zaitsu, M. Takenaka and Y. Nakano, "Monolithic InP 100-port photonic switch," in proceedings of 36th European Conference and Exhibition on Optical Communication (ECOC2010), PD1.5. (2010).
13. M.-J. Kwack, T. Tanemura, A. Higo, and Y. Nakano, "Monolithic InP strictly non-blocking 8×8 switch for high-speed WDM optical interconnection," Opt. Express **20**(27), 28734–28741 (2012).
14. I. M. Soganci, T. Tanemura, and Y. Nakano, "Integrated phased-array switches for large-scale photonic routing on chip," Laser Photonics Rev. **6**(4), 549–563 (2012).
15. F. Aflatouni, B. Abiri, A. Rekhi, and A. Hajimiri, "Nanophotonic projection system," Opt. Express **23**(16), 21012–21022 (2015).
16. M. C. Shin, A. Mohanty, K. Watson, G. R. Bhatt, C. T. Phare, S. A. Miller, M. Zadka, B. S. Lee, X. Ji, I. Datta, and M. Lipson, "Chip-scale blue light phased array," Opt. Lett. **45**(7), 1934–1937 (2020).
17. T. Fukui, Y. Kohno, R. Tang, Y. Nakano, and T. Tanemura, "Single-pixel imaging using multimode fiber and silicon photonic phased array," J. Light. Technol. **39**(3), 839-844 (2021).
18. H. Abediasl and H. Hashemi, "Monolithic optical phased-array transceiver in a standard SOI CMOS process," Opt. Express **23**(5), 6509–6519 (2015).
19. S. W. Chung, H. Abediasl, and H. Hashemi, "A monolithically integrated large-scale optical phased array in silicon-on-insulator CMOS," IEEE J. Solid-State Circuits **53**(1), 275–296 (2018).
20. R. Fatemi, A. Khachaturian, and A. Hajimiri, "A nonuniform sparse 2-D Large-FOV optical phased array with a Low-Power PWM drive," IEEE J. Solid-State Circuits **54**(5), 1200–1215 (2019).
21. W. Guo, P. R. A. Binetti, C. Althouse, M. L. Mašanovic, H. P. M. M. Ambrosius, L. A. Johansson, and L. A. Coldren, "Two-Dimensional optical beam steering with InP-Based photonic integrated circuits," IEEE J. Sel. Top. Quantum Electron. **19**(4), 6100212 (2013).
22. W. Guo, P. R. A. Binetti, M. L. Masanovic, L. A. Johansson, and L. A. Coldren, "Large-scale InP photonic integrated circuit packaged with ball grid array for 2D optical beam steering," in 2013 IEEE Photonics Conference, pp. 651–652.
23. J. C. Hulme, J. K. Doylend, M. J. R. Heck, J. D. Peters, M. L. Davenport, J. T. Bovington, L. A. Coldren, and J. E. Bowers, "Fully integrated hybrid silicon two dimensional beam scanner," Opt. Express **23**(5), 5861–5874 (2015).
24. W. Xie, T. Komljenovic, J. Huang, M. Tran, M. Davenport, A. Torres, P. Pintus, and J. Bowers, "Heterogeneous silicon photonics sensing for autonomous cars," Opt. Express **27**(3), 3642–3663 (2019).
25. S. Tan, J. Liu, Y. Liu, H. Li, Q. Lu, and W. Guo, "Two-dimensional beam steering based on LNOI optical phased array," in Conference on Lasers and Electro-Optics, (Optical Society of America, 2020), SM2M.1.
26. Y. Hirano, Y. Miyamoto, M. Miura, Y. Motoyama, K. Machida, T. Yamada, A. Otomo, and H. Kikuchi, "High-speed optical-beam scanning by an optical phased array using electro-optic polymer waveguides," IEEE Photonics J. **12**(2), 6600807 (2020).
27. D. R. Wight, J. M. Heaton, B. T. Hughes, J. C. H. Birbeck, K. P. Hilton, and D. J. Taylor, "Novel phased array optical scanning device implemented using GaAs/AlGaAs technology," Appl. Phys. Lett. **59**(8), 899–901 (1991).
28. N. A. Tyler, D. Fowler, S. Malhouitre, S. Garcia, P. Grosse, W. Rabaud, and B. Szelag, "SiN integrated optical phased arrays for two-dimensional beam steering at a single near-infrared wavelength," Opt. Express **27**(4), 5851–5858 (2019).
29. M. Prost, Y. Ling, S. Cakmakyapan, Y. Zhang, K. Zhang, J. Hu, Y. Zhang, and S. J. B. Yoo, "Solid-State MWIR beam steering using optical phased array on Germanium-Silicon photonic platform," IEEE Photonics J. **11**(6), 6603909 (2019).

30. J. Midkiff, K. M. Yoo, J.-D. Shin, H. Dalir, M. Teimourpour, and R. T. Chen, "Optical phased array beam steering in the mid-infrared on an InP-based platform," Optica **7**(11), 1544–1547 (2020).
31. C. V. Poulton, M. J. Byrd, B. Moss, E. Timurdogan, R. Millman, and M. R. Watts, "8192-element optical phased array with 100° steering range and Flip-Chip CMOS," in Conference on Lasers and Electro-Optics, (Optical Society of America, 2020), JTh4A.3.
32. L. Chrostowski, X. Wang, J. Flueckiger, Y. Wu, Y. Wang, and S. T. Fard, "Impact of fabrication non-uniformity on chip-scale silicon photonic integrated circuits," in Optical Fiber Communication Conference, OSA Technical Digest (online) (Optical Society of America, 2014), paper Th2A.37.
33. Y. Yang, Y. Ma, H. Guan, Y. Liu, S. Danziger, S. Ocheltree, K. Bergman, T. Baehr-Jones, and M. Hochberg, "Phase coherence length in silicon photonic platform," Opt. Express **23**(13), 16890–16902 (2015).
34. M. Sun and J. Zhang, "Single-Pixel Imaging and Its Application in Three-Dimensional Reconstruction: A Brief Review," Sensors **19**(3), 732(2019).
35. K. Komatsu, Y. Ozeki, Y. Nakano, and T. Tanemura, "Ghost imaging using integrated optical phased array," in Optical Fiber Communication Conference, (Optical Society of America, 2017), Th3H.4 (2017).
36. L. Li, W. Chen, X. Zhao, and M. Sun, "Fast optical phased array calibration technique for random phase modulation LiDAR," IEEE Photonics J. **11**(1), 6900410 (2019).
37. Y. Kohno, K. Komatsu, R. Tang, Y. Ozeki, Y. Nakano, and T. Tanemura, "Ghost imaging using a large-scale silicon photonic phased array chip," Opt. Express **27**(3), 3817–3823 (2019).
38. J. Ø. Kjellman, M. Prost, A. Marinins, H. K. Tyagi, T. D. Kongnyuy, S. Kerman, B. Troia, B. Figeys, S. Dwivedi, M. S. Dahlem, P. Soussan, X. Rottenberg, and R. Jansen, "Silicon photonic phase interrogators for on-chip calibration of optical phased arrays," Proc. SPIE **11283**, 112830X (2020).
39. J. Shim, J.-B. You, H.-W. Rhee, H. Yoon, S.-H. Kim, K. Yu, and H.-H. Park, "On-chip monitoring of far-field patterns using a planar diffractor in a silicon-based optical phased array," Opt. Lett. **45**(21), 6058–6061 (2020).
40. S. Chung, M. Nakai, S. Idres, Y. Ni and H. Hashemi, "Optical phased-array FMCW LiDAR with on-chip calibration," in 2021 IEEE International Solid- State Circuits Conference (ISSCC), pp. 286-288.
41. D. N. Hutchison, J. Sun, J. K. Doylend, R. Kumar, J. Heck, W. Kim, C. T. Phare, A. Feshali, and H. Rong, "High-resolution aliasing-free optical beam steering," Optica **3**(8), 887–890 (2016).
42. K. Sayyah, O. Efimov, P. Patterson, J. Schaffner, C. White, J.-F. Seurin, G. Xu, and A. Miglo, "Two-dimensional pseudo-random optical phased array based on tandem optical injection locking of vertical cavity surface emitting lasers," Opt. Express **23**(15), 19405–19416 (2015).
43. T. Komljenovic, R. Helkey, L. Coldren, and J. E. Bowers, "Sparse aperiodic arrays for optical beam forming and LIDAR," Opt. Express **25**(3), 2511–2528 (2017).
44. B. Yang, H. Chen, S. Yang, and M. Chen, "An improved aperiodic OPA design based on large antenna spacing," Opt. Commun. **475**, 125852 (2020).
45. Indeed, $R_{uu}(\Delta)$ corresponds to the term $|U(\xi)|^2$ of $I(\xi)$ in Eq. (5) and thus only influences the envelope function of the FFP.
46. Such assumption is justified since once we obtain a sharp focused beam at $\xi_0 = (0,0)$, it could be steered to an arbitrary 2D angle by satisfying Eq. (6).
47. A. Moffet, "Minimum-redundancy linear arrays," IEEE Trans. Antennas Propag. **16**(2), 172–175 (1968).
48. J. P. Costas, "A study of a class of detection waveforms having nearly ideal range—Doppler ambiguity properties," Proc. IEEE **72**(8), 996-1009 (1984).
49. W. C. Babcock, "Intermodulation interference in radio systems frequency of occurrence and control by channel selection," in The Bell Syst. Tech. J. **32**(1), 63-73 (1953).
50. S.W. Golomb, "Algebraic constructions for Costas arrays," J. Cominatorial Theory Series A **37**(1), 13-31 (1984).
51. J. K. Beard, J. C. Russo, K. G. Erickson, M. C. Monteleone and M. T. Wright, "Costas array generation and search methodology," IEEE Trans. Aerosp. Electron. Syst **43**(2), 522-538 (2007).
52. A. Dollas, W. T. Rankin and D. McCracken, "A new algorithm for Golomb ruler derivation and proof of the 19 mark ruler," IEEE Trans. Inf. Theory **44**(1), 379-382 (1998).
53. K. Drakakis, "A review of the available construction methods for Golomb rulers," Adv. Math. Comm. **3**(3), 235-250 (2009).
54. T. Fukui, Y. Nakano, and T. Tanemura, "Resolution limit of single-pixel speckle imaging using multimode fiber and optical phased array," J. Opt. Soc. Am. B **38**(2), 379-386 (2021).
55. S. Mano and T. Katagi, "A method for measuring amplitude and phase of each radiating element of a phased array antenna," IEICE Trans. **J65-B** (5), 555–560 (1982).
56. Q. Zhang, L. Zhang, Z. Li, W. Wu, G. Wang, X. Sun, and W. Zhang, "An antenna array initial condition calibration method for integrated optical phased array," arXiv:1902.06203 (2019).